\PassOptionsToPackage{numbers,sort&compress}{natbib}
\documentclass[10pt]{article}
\usepackage[margin=1in]{geometry}
\usepackage{lipsum} 
\usepackage{ragged2e} 
\usepackage{mathpazo}
\usepackage{fancyhdr}
\usepackage{lastpage}
\usepackage{amsmath,amssymb}
\usepackage{authblk}  
\usepackage{appendix} 
\usepackage{graphicx}
\usepackage[numbers]{natbib}  
\usepackage{soul}
\usepackage{xcolor}   
\usepackage{enumitem}
\usepackage{rotating}
\usepackage{makecell}
\usepackage{float}    
\usepackage[numbers,sort&compress]{natbib}
\usepackage{indentfirst}
\usepackage{xcolor}
\usepackage[colorlinks=true,
            linkcolor=mylinkcolor,
            citecolor=mylinkcolor,
            urlcolor=mylinkcolor]{hyperref}

\definecolor{mylinkcolor}{RGB}{50,120,150}

\date{}

\newenvironment{keywords}{
  \par\vspace{0.5em}\noindent
  \textbf{Keywords: }
  \itshape
}{\par}

\begin{document}

\begin{center}
    {\huge \textbf{Pseudo-spectral model of elastic-wave propagation through toothed-whale head anatomy, and implications for biosonar}}
\end{center}
\vspace{0.5em} 
\hrule height 1pt 
\vspace{1em} 
\noindent
\begin{minipage}[t]{0.38\textwidth}
  \raggedright

  \textbf{Fawad Ali}\textsuperscript{a,*} \\ 
  \textbf{Carlos García A.}\textsuperscript{a} \\ 
  \textbf{Aida Hejazi-Nooghabi}\textsuperscript{b} \\ 
  \textbf{Lapo Boschi}\textsuperscript{a} \\[0.8em]

  \textit{\textsuperscript{(a)} Dipartimento di Geoscienze,  Università degli Studi di Padova, 35131, Padova, Italy} \\[0.5em]

  \textit{\textsuperscript{(b)} Institute of Geophysics, Centre for Earth System Research and Sustainability (CEN),  Universität Hamburg, 20146, Hamburg, Germany} \\[0.5em]

  \textsuperscript{*} {\small {Email: \texttt{fawad.ali@unipd.it}}}
\end{minipage}%
\hfill
\begin{minipage}[t]{0.58\textwidth}
\raggedright
{\large \textbf{Abstract}} \\[0.65em]
\justifying
\noindent
The sound-localization and, in particular, biosonar system of toothed whales is exceptionally performant. How this is achieved is not clear, given that: (i) toothed whales have no pinnae; (ii) while their auditory pathways have been studied in detail, no specific feature apparently replacing the pinna has been identified. In this study, we employ a pseudo-spectral time domain (PSTD) numerical scheme to model three-dimensional elastic wave propagation through a toothed-whale head including soft tissues. Computed tomography (CT) scans were utilized to build a three-dimensional velocity-density model of the specimen's head, parametrized on a high-resolution $1.11$ mm voxel grid. We first validate our wave propagation solver, identifying a range of frequencies and spatial scale lengths where the PSTD scheme captures the complexities of elastic wave propagation through toothed-whale anatomy. We next focus on the toothed whale's ability to locate sources on the median plane, where the role of anatomy is crucial. A 45 kHz central frequency burst (dolphin-like click) was modeled and directed at elevation angles from $-90^\circ$ to $+90^\circ$ in $5^\circ$ steps along the midsagittal plane. We find that the incoming sound can be localized, via correlation, from the reverberated portion of the time-domain waveforms recorded at the tympano-periotic complex locations.   
\end{minipage}

\vspace{1em}
\begin{keywords}
Toothed Whale Echolocation, Pseudo-Spectral Method, Numerical Modeling of wave propagation, Auditory system
\end{keywords}
\vspace{0.5em} 
\hrule height 1pt 
\vspace{1em} 
\section{\label{sec:1} Introduction}
Terrestrial mammals localize sound sources in a high impedance mismatch medium, i.e., the animal anatomy and the surrounding air\citep{grothe2010mechanisms}. In \textit{marine} mammals the impedance mismatch is minimal, as the density of biological tissues is similar to that of water, allowing efficient transmission of acoustic energy into cetaceans' bodies as both compressional and shear waves\citep{ladich2017acoustic}. Perhaps due to the relative simplicity of acoustic wave propagation, the auditory systems of many terrestrial species are better understood than those of cetaceans\citep{carlini2024auditory, jones2007bat, mooney2012hearing}. Cetaceans' audition has been studied through anatomical dissection and behavioral audiometry, revealing specialized adaptations for sound localization in marine environments\citep{de2023neuroanatomy}. It has been established that toothed whales receive sound primarily through their lower jaws, as initially proposed by Norris (1968). This ``acoustic window'' consists of lipid-filled structures along the lower jaw that efficiently conduct sound to the inner ear bones, which are acoustically isolated from surrounding tissues\citep{norris1964some, houser1999classification}.

Additional or alternative routes to Norris’ proposed acoustic window include the propagation through the throat region\citep{cranford2010new}; bone conduction, which plays a key role for baleen whales\citep{nummela2007sound, yamato2012auditory, cranford2015fin, morris2025experimental}, might also be relevant; mental foramina---the paired openings in the lower jaw that house nerves and blood vessels---have been claimed to serve as potential acoustic entry points contributing to directional hearing\citep{ryabov2010role}; air-filled sinuses that surround the tympano-periotic complex (TPC) apparently help isolate the ears from the rest of the skull and from one another\citep{cranford2008acoustic}. (If the inner ears were not acoustically isolated, both ears would receive a very complex signal, multiply reverberated by the skull and surrounding tissues, and presumably very hard to interpret in terms of source localization.)  The different sound-propagation pathways that are now being debated are not necessarily mutually exclusive. Rather, they could play different roles in a complex mechanism, in which the specific route of sound conduction is correlated with the frequency of the incoming signal\citep{ketten1997structure}. 

Mammals localize sound based both on the interaural time differences (ITDs) and interaural level differences (ILDs) between left and right ears, and on the filtering effect introduced by the pinnae, which control, to a large extent, their head-related transfer functions (HRTFs). Most terrestrial mammals can precisely identify the azimuth of incoming sound but are less accurate at determining elevation\citep{middlebrooks2015sound}. On the other hand, behavioural experiments\citep{renaud1976sound} show that the biosonar of toothed whales is approximately equally sensitive to changes in elevation vs. azimuth of the target. Estimates of the so-called minimum audible angle, or MAA (the smallest angular separation between two sound sources that a subject can distinguish as being in different places) of, e.g., the bottlenose dolphin are as small as $<1^{\circ}$ both horizontally and vertically. (In comparison, the MAA of humans is also about $1^{\circ}$ horizontally, but much worse--several degrees--vertically\citep{van2016auditory}.) How such accuracy is achieved is unclear, given that: (i) toothed whales have no pinnae and no specific feature replacing the pinnae has been identified; (ii) we currently have little knowledge of how their brain processes the information it receives from the ears\citep{de2023neuroanatomy}. It has been inferred that their localization mechanism must work in a profoundly different way than those of other species\citep{nachtigall2016biosonar}, but it is currently not known if certain anatomical features have evolved to functionally replace the pinnae. It is also not known if a localization algorithm within the brain exists, capable of processing and interpreting more complex information than pinnae-generated spectral peaks and  notches\citep{nachtigall2016biosonar}. 

In this study, we model elastic-wave propagation through the dolphin anatomy, with the specific goal of understanding whether anatomy filters auditory information in such a way as to facilitate the task of sound localization, much like the pinnae of terrestrial mammals. Since the role of the pinnae is limited to median-plane (the vertical plane that divides the subject’s head into left and right halves) localization, and azimuthal localization is presumably achieved by evaluating interaural source or level differences, we focus here on signals generated by sources on the median plane only. Similar to Hejazi-Nooghabi et al. (2021)\citep{nooghabi2021contribution}, we employ the Pseudo-Spectral Time Domain (PSTD) numerical model, further extending it to include soft tissues. In the context of numerical schemes, several Finite Element Method (FEM)-based models have already shown promising results in understanding sound propagation through the anatomical components of the cetacean’s head, particularly dolphins\citep{wei2016role, song2016inducing, zhang2017directional, wei2018numerical, wei2018finite, wei2020modeling, wei2022distinctive}. However, the majority of FEM-based studies are either limited to the frequency domain, or, if in the time domain, to two-dimensional (2D) analogues of dolphin anatomy, with the exception of a recent three-dimensional (3D) transient FEM model of the Risso’s dolphin (Grampus griseus) limited to frequencies  $<$ 40 kHz\citep{wei2024validated}. The PSTD scheme offers a memory-efficient solution that requires only 2 points per wavelength as per the Nyquist-Shannon sampling theorem\citep{oppenheim1997signals}. This study is the first attempt to apply the PSTD method to elastic wave propagation in a realistic toothed-whale anatomy model (including both bone and soft tissues), evaluating its limitations and potential for future applications. 

This paper is structured into three main parts. The first part describes the numerical modelling setup, including validation and assessment of its accuracy. The second part illustrates our numerical simulations, focusing on the dolphin’s localization of sound sources within the median plane. Finally, the third part interprets the results, draws conclusions, and suggests directions for future work.

\begin{figure}[h]
\centering
\includegraphics[width=4.5in]{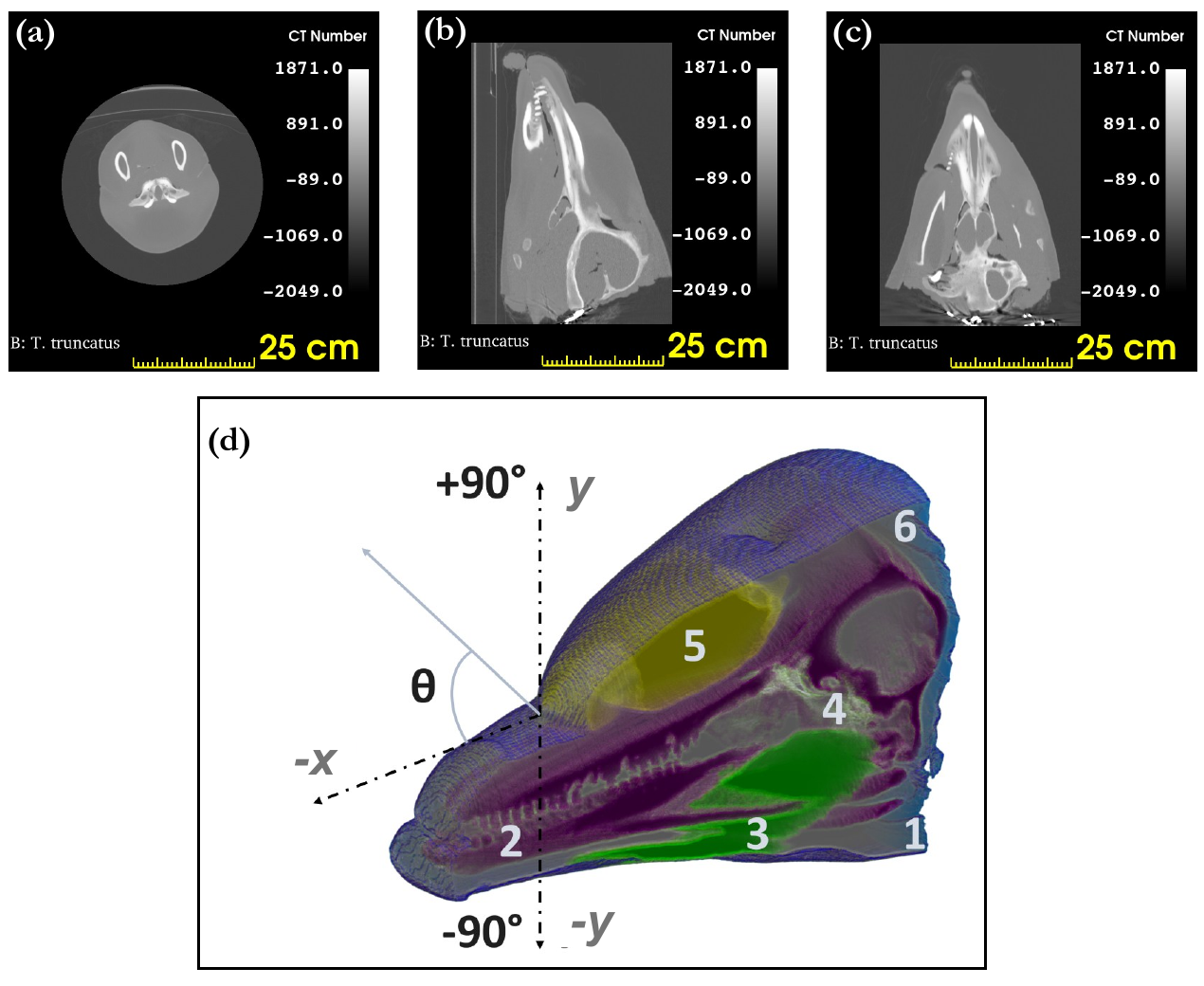}
\caption{\label{fig:FIG1}{Computed tomography (CT) scans of the dolphin’s head immersed in water, shown as grayscale CT slices in (a) axial, (b) sagittal, and (c) coronal planes. (d) Reconstructed model of the dolphin’s head in MATLAB, with labeled anatomical components: epidermis(1), bone(2), mandibular fat(3), air(4), melon(5), and connective tissue(6).}}
\end{figure}

\section{\label{sec:2}	NUMERICAL MODEL SETUP}
High-resolution images of a deceased toothed-whale (Tursiops truncatus) head, acquired through a CT scanner set to 120 kV and 150 mAs, were provided to us by colleagues at the Department of Comparative Biomedicine and Food Science, University of Padua. The image data in DICOM format were imported into 3D Slicer, an open-source software platform for medical image visualization and analysis. Using CT number thresholds (measured in Hounsfield units, or HU), the tissue volumes were cleaned, segmented, and categorized into distinct anatomical structures such as mandibular fats, the melon, connective tissues, and bones, as illustrated in Figs.~\ref{fig:FIG1}a–\ref{fig:FIG1}c. Each segmented tissue was then exported as a Nearly Raw Raster Data (NRRD) file, which was translated to 3D MATLAB-compatible data by means of Python libraries such as Pynrrd, NumPy, etc. The head model was oriented so that its median plane is approximately parallel to the x-axis of our reference frame (Fig.~\ref{fig:FIG1}d). We transform the x-ray attenuation coefficient in HU to mass density  via the formulae,
\begin{subequations}
\begin{eqnarray}
\rho_{\mathrm{tissue}} &=& 0.02988\,\mathrm{HU}^2 + 0.4938\,\mathrm{HU} + 954.794, \label{eq:rho_tissue} \\[4pt]
\rho_{\mathrm{bone}}   &=& 11.7\,\mathrm{HU} - 10944.75, \label{eq:rho_bone} \\[4pt]
c_{p,\mathrm{tissue}}  &=&
\begin{cases}
2.303\,\rho - 809.77, & \rho < 1000, \\[4pt]
1.76836\,\rho - 282.292, & \rho \ge 1000,
\end{cases} \label{eq:cP_tissue}
\end{eqnarray}
\end{subequations}
\begin{table}[ht]
\caption{\label{tab:table1}Assumed values of the elastic parameters in our model}
\renewcommand{\arraystretch}{1.3} 
\setlength{\tabcolsep}{4pt} 
\begin{tabular}{c c c c c c} 
\hline\hline
\text{Parts} & 
\begin{tabular}[c]{@{}c@{}}Hounsfield\\ Number (HU)\end{tabular} & 
\begin{tabular}[c]{@{}c@{}}Averaged\\ Density ($\rho$)\\ {[}kg·m$^{-3}${]}\end{tabular} & 
\begin{tabular}[c]{@{}c@{}}Averaged\\ Compressional\\ Wave Speed ($c_p$)\\ {[}m·s$^{-1}${]}\end{tabular} & 
\begin{tabular}[c]{@{}c@{}}Poisson\\ Ratio ($\nu$)\end{tabular} & 
\begin{tabular}[c]{@{}c@{}}Averaged\\ Shear Wave Speed\\ {[}m·s$^{-1}${]}\\ $c_p\sqrt{\dfrac{1-2\nu}{2(1-\nu)}}$\end{tabular} \\ 
\hline
Bone & 150 to 2005 & 1662 & 3050.8 & 0.3 & 1630.7 \\ \hline
Connective tissues & 20 to 90 & 1018 & 1517.9 & 0.49 & 227.9 \\ \hline
Epidermis & 20 to 140 & 1106.5 & 1674.4 & 0.49 & 233.9 \\ \hline
Mandibular fats & -70 to 10 & 996.5 & 1485.2 & 0.49 & 207.9 \\ \hline
Melon & -104 to -64 & 956.9 & 1394 & 0.49 & 195.2 \\ 
\hline\hline
\end{tabular}
\end{table}

\noindent
where $\rho_{\mathrm{tissue}}$, $\rho_{\mathrm{bone}}$ and $c_{p,\mathrm{tissue}}$ denote the density of tissues and bones (in~kg$\cdot$m$^{-3}$) and the compressional wave speed of tissues (in~m$\cdot$s$^{-1}$), respectively. Since HU to density conversion factor were not provided and as the study's primary objective is to evaluate the numerical scheme's performance on soft tissues, Eqs.~(\ref{eq:rho_tissue}), (\ref{eq:rho_bone}), and (\ref{eq:cP_tissue}) are heuristically derived based on the existing literature\citep{nooghabi2021contribution, goss1978comprehensive, soldevilla2005cuvier, emelianov2006synergy, wei2023does}.  The resulting model is illustrated in Fig.~\ref{fig:FIG1}d. For the sake of simplicity, we averaged the values of elastic parameters within individual anatomical features, each of which is then defined by one value of $\rho$, $c_{p}$, and $\upsilon$ as per Table~\ref{tab:table1}. Importantly, we decided to neglect viscoelasticity due to the insufficient data available in the literature on wave attenuation in melon, fats, and other connective tissues. A lossless model enhances the wave field’s reverberated ``coda,'' i.e., the small-amplitude slowly decaying wave field generated by multiple scattering after the arrival of the main signal, which may be the most important effect of the anatomy. To account for shear wave speed, the Poisson ratio was carefully chosen for both solid and soft tissue. The Poisson ratio of bones is set to 0.3, assuming them to be purely elastic, whereas for soft tissues such as mandibular fats, melon, and connective tissues, the Poisson ratio was kept as close as possible to 0.5\citep{vannah1993modelling}.
\subsection{\label{subsec:1:1}Elastic wave solver (k-Wave Toolbox)}
The simulation domain comprises the head of the dolphin and the water in which it is immersed, forming a three-dimensional, 590 mm by 416 mm by 416 mm rectangular box. The computational grid is reproduced in MATLAB utilizing the k-Wave PSTD software package\citep{treeby2014modelling}. This method employs Fourier basis functions to expand the spatial derivatives, resulting in an ordinary differential equation of the form
\begin{equation}
\frac{\partial f(\mathbf{r})}{\partial x} = \mathcal{F}^{-1} \left\{ i k_{x} \, \mathcal{F}\big(f\big) \right\},
\label{eq4}
\end{equation}
where $x$, $y$, $z$ are the Cartesian coordinates, ${\bf k}=(k_x, k_y, k_z)$ is the wave vector, $i$ is the imaginary unit and $\mathcal{F}$ denotes the Fourier transform operator. This approach only supports periodic boundaries, where reflections occur on the opposite sides of the domain, and does not accommodate other types of boundaries, such as absorbing or traction/free surfaces\citep{igel2017computational}. To prevent unwanted reflections, a Split-field Perfectly Matched Layer (PML) is applied to each boundary\citep{berenger1994perfectly, berenger1996three}. The Fourier basis is preferred over Chebyshev polynomials (which can handle non-periodic boundaries) due to its high accuracy\citep{igel2017computational}, absence of the Runge phenomenon\citep{boyd2001chebyshev}, and a significantly lower computational cost, thanks to relaxed time-stepping\citep{gottlieb1991cfl} and the Fast Fourier Transform (FFT) algorithm\citep{chu1999inside}.  The time-marching algorithm is implemented by a non-standard finite-difference scheme\citep{mickens2020nonstandard}. The physics is based on the solution of the linear elastic-acoustic model, 
\begin{subequations}
\begin{equation}
\frac{\partial \sigma_{ij}}{\partial t}
= \lambda \, \delta_{ij} \frac{\partial v_{k}}{\partial x_{k}}
+ \mu \left( \frac{\partial v_{i}}{\partial x_{j}} + \frac{\partial v_{j}}{\partial x_{i}} \right),
\label{eq5}
\end{equation}
\begin{equation}
\rho\frac{ \partial v_{i}}{\partial t}
=  \frac{\partial \sigma_{ij}}{\partial x_{j}},
\label{eq6}
\end{equation}
\end{subequations}
where $\sigma_{ij}$ denotes the  $i, j$ component of the stress tensor, $v_i$ the $i$ component of the particle velocity, and $\rho$ the density of the medium. $\lambda$ and $\mu$ are the Lam\'e coefficients, which are related to the shear wave speed $c_s$ and the compressional wave speed $c_p$ through
\begin{equation}
\mu = c_s^2 \rho, 
\end{equation}
\begin{equation}
\mu + 2\lambda = c_p^2 \rho. 
\end{equation}
The elastic constitutive relation holds, i.e.,
\begin{equation}
\sigma_{ij} = \lambda \delta_{ij} \, \varepsilon_{kk} + 2\mu \varepsilon_{ij}, 
\end{equation}
\begin{equation}
\varepsilon_{ij} = \frac{1}{2} \left( \frac{\partial u_i}{\partial x_j} + \frac{\partial u_j}{\partial x_i} \right), 
\end{equation}
where $\delta_{ij}$ is the Kronecker delta and $u_i$ is the $i$-th component of displacement. The acoustic pressure in fluid media is given by
\begin{equation}
p(x,y,z,t) = -\frac{\mathrm{Tr}(\sigma_{ij})}{3}, 
\end{equation}
where $\mathrm{Tr}$ denotes the trace of a tensor.

\begin{figure}[h]
\centering
\includegraphics[width=3.5in]{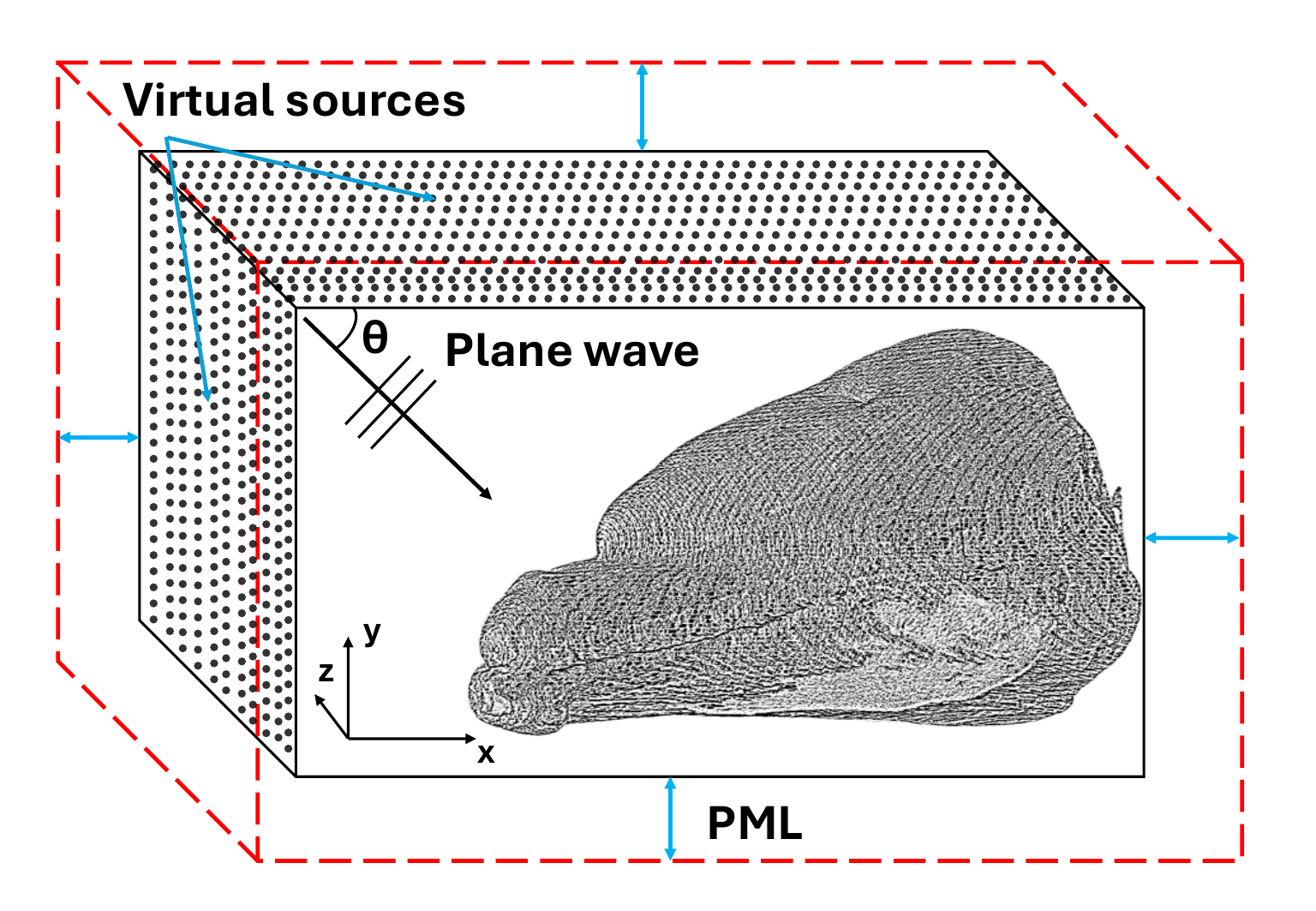}
\caption{\label{fig:FIG2}{Numerical simulation setup. Black dots on the top and left sides of the smaller box mark the virtual point sources used for generating the incoming plane wave. The red dashed lines the full extent of the box including the PML layers.}}
\end{figure}

In this study, we limit ourselves to relatively far sources, whose generated sound field can be approximated by a plane wave. To keep computational cost to a minimum, we deploy an array of virtual sources along the boundaries of our numerical simulation domain; we use analytical formulae to calculate when the wavefront hits each of those points and have our source ``fire'' accordingly. Since we focus on determining source elevation, we model sources within the median plane only. The simulation domain is a hexahedral box with vertical faces aligned parallel and perpendicular to the median plane. The dolphin head is aligned to face the $xz$ plane, and the $x$ direction follows the beak direction, as shown in Fig.~\ref{fig:FIG2}. This ensures all incoming wavefronts are perpendicular to the box’s vertical sides.  Said $\theta$ the angle that the propagation direction forms with the horizontal, the time $t_V$ for the plane wave to reach a ``source'' on the horizontal plane is
\begin{subequations}
\begin{eqnarray}
t_V = -\frac{N d}{c_0} \, \sin\theta. 
\end{eqnarray}
While the time when it reaches a ``source” on the vertical plane is given by
\begin{eqnarray}
t_H = -\frac{N d}{c_{p0} }\, \cos\theta, 
\end{eqnarray}
\end{subequations}
where $N$ denotes the number of point sources along one edge of the inner box in Fig.~\ref{fig:FIG2}: 372 in the vertical direction and 530 in the horizontal one, $d$ the distance between two neighbouring sources, and $c_{p0}$ the compressional wave speed in water. The pressure $p(t)$ is given by
\begin{equation}
p(t) = \cos\theta S_V (t + t_V) + \sin\theta S_H (t + t_H),
\label{eq11}
\end{equation}
where $S_V$ denotes the pressure field arising from sources on the vertical boundary of the box and $S_H$ indicates the contribution of sources on the horizontal boundaries. The cosine and sine factors in Eq.~(\ref{eq11}) normalize the amplitude of the plane wave regardless of its angle of arrival.
\subsection{Validation of the numerical scheme}
\begin{figure}[h]
\centering
\includegraphics[width=4.5in]{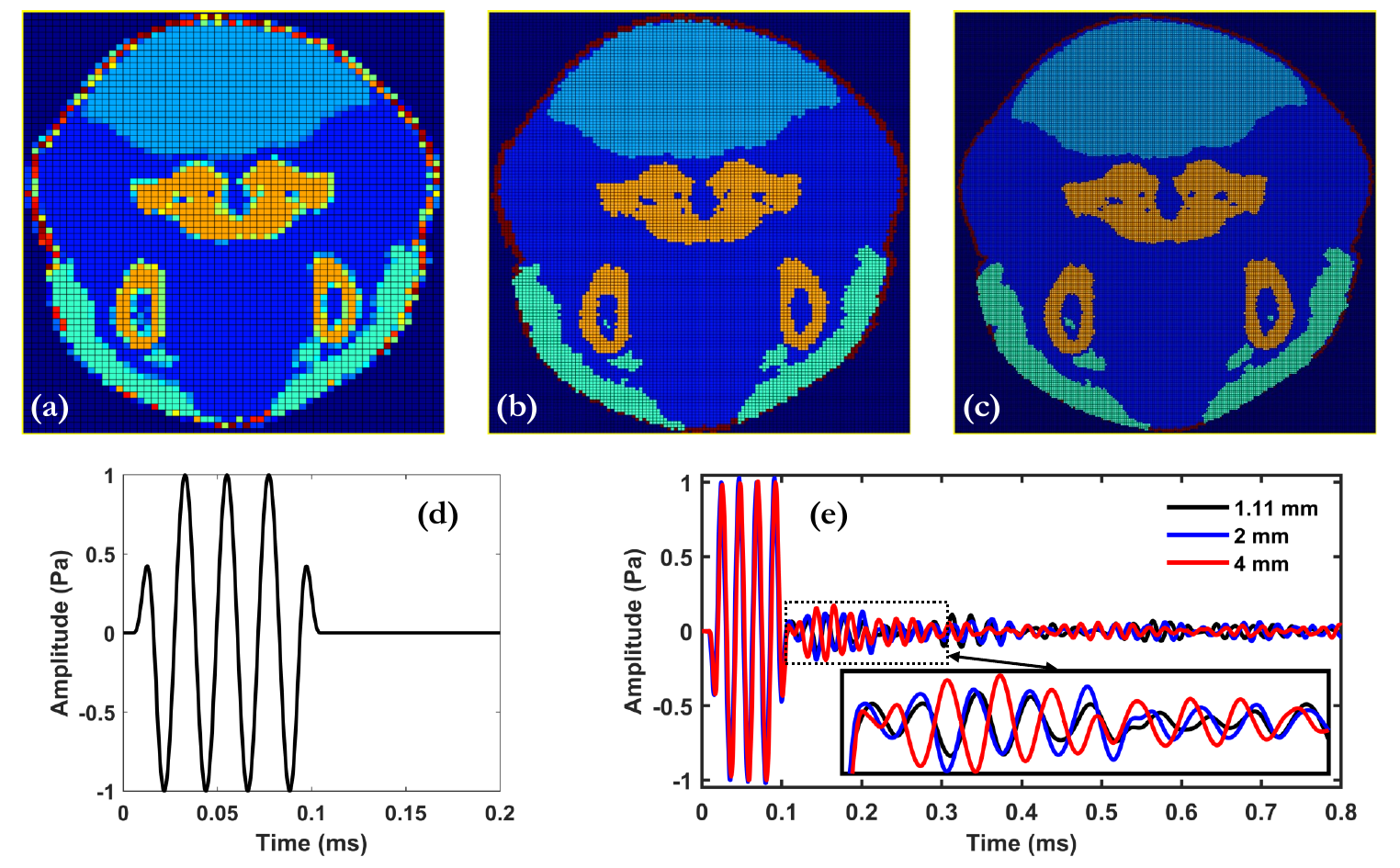}
\caption{\label{fig:FIG3}{Visualization of the voxel grid in the axial view of the dolphin’s head CT scan after down sampling to (a) 4 mm, (b) 2 mm, and (c) 1.11 mm voxel resolutions. (d) Input sinusoidal burst source used in the simulations. (e) Results of the convergence test, showing a comparison of recorded waveforms corresponding to (a), (b), and (c).}}
\end{figure}

The accuracy of our PSTD solver clearly depends on the resolution of the regular grid used to parameterize the medium. A convergence test was conducted by down-sampling the CT model to various resolutions: $1.11$ mm, 2 mm, and 4 mm. Visual inspection of Figs.~\ref{fig:FIG3}a–\ref{fig:FIG3}c, suggests that the 4 mm voxel model does not adequately capture the shape of interfaces. The 2-mm and $1.11$-mm models substantially reduce pixelization. We assume that 10 points per wavelength, i.e., 5 times higher than the Nyquist-Shannon criterion are enough to properly capture wave propagation through the model. This means that the 4-mm model can resolve up to $37.5$ kHz, with the possibility of errors arising from distorted interfaces in the geometric model, the 2-mm model up to 75 kHz, and the $1.11$-mm model up to 135 kHz. In the following, we shall use the 1.11 mm parameterization with maximum wave frequency of about 90 kHz.

We further tested these models by considering a $590 \times 416 \times 416$~mm simulation box with 20~mm thick perfectly matched layer (PML) boundaries. A plane-wave sinusoidal burst with a central frequency of 45~kHz, windowed with a Tukey window (Fig.~\ref{fig:FIG3}d), was injected into the model as described in more detail in section~\ref{subsec:1:1}. A virtual receiver was positioned along the midsagittal plane of the specimen’s head, near the jaw tip, with coordinates $(x, y, z) = (55, 208, 208)$ mm. The pressure amplitudes recorded are compared across the three models, as shown in Fig.~\ref{fig:FIG3}e. All models demonstrated satisfactory direct wave propagation, but a distinct difference in waveform emerged upon the onset of reverberation. Specifically, between $0.1$ and $0.3$ milliseconds, the $4$-mm is out of phase with respect to both the $2$-mm and $1.11$-mm models, which instead are quite similar, albeit with some discrepancies in amplitude. The phase discrepancy arises from pixelization, causing a temporal shift in the waveform (phase error in time shift = voxel size/wave speed), as the 4-mm model is unable to capture curved surfaces smaller than its own voxel size, thereby accumulating error, particularly at solid-fluid discontinuities\citep{van2002finite}. Conversely, the 2-mm and $1.11$-mm models result in similar waveforms, both in phase and amplitude, at least up to $0.8$ ms.  

\begin{figure}[h]
\centering
\includegraphics[width=4.5in]{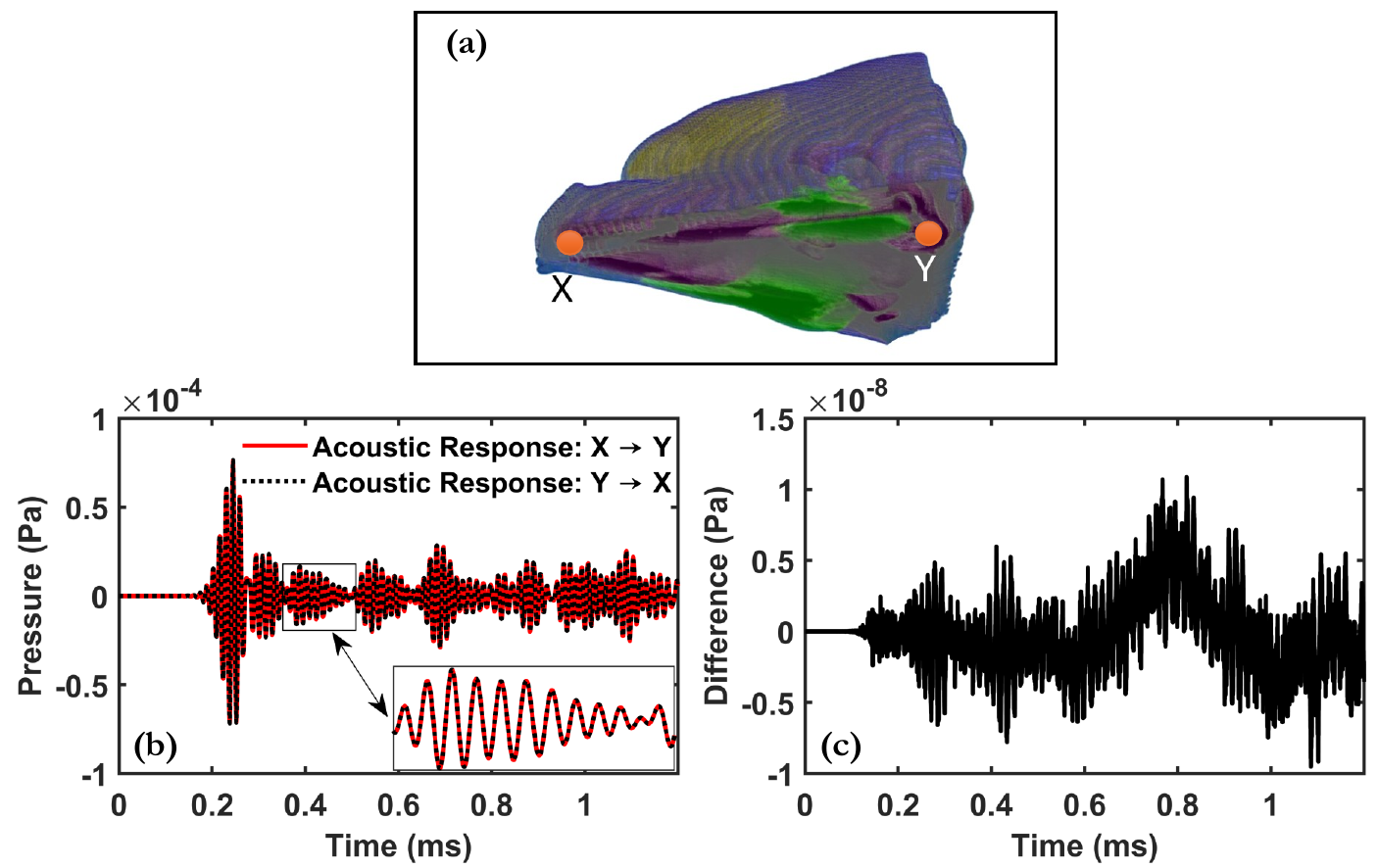}
\caption{\label{fig:FIG4}{Validation of the numerical model. (a) Schematic illustrating the positions of points X, and Y located in the specimen's head, at the tip of the jaw, and at the TPC, respectively. (b) Comparison of pressure waveforms recorded at positions Y and X, respectively. (c) Error difference calculated by subtracting the waveforms of the forward and reverse simulation.}}
\end{figure}
Next, we utilize the spatial reciprocity theorem to numerically validate the current model\citep{aki2002quantitative}. The spatial resolution of $1.11$ mm is used, and, again, a plane-wave sinusoidal burst plane wave with a central frequency of 45 kHz. The PML is set up as above. The point source was placed at the jaw tip, (position $X$ in Fig.~\ref{fig:FIG4}a) at coordinates $(x,y,z) = (49, 188, 248)$ mm, and the sensor at the TPC, (position $Y$ in Fig.~\ref{fig:FIG4}a) with coordinates $(x,y,z) = (400, 92, 198)$ mm. Initially, a point source was deployed at $X$ and recorded at $Y$. Next, the positions of source and receiver were swapped, and the signal was recorded at $X$. Upon comparison, it was observed that the  relative difference between the two reciprocal recordings was of the order of $10^{-4}$, as shown in Figs.~\ref{fig:FIG4}b and~\ref{fig:FIG4}c. A similar test was conducted with both points $X$ and $Y$ in the fluid domain, resulting in equivalent order-of-magnitude discrepancies.
\subsection{Potential issues at fluid-solid interfaces}
While results in the previous section indicate that our PSTD approach is adequate to model elastic wave propagation in a dolphin's head, including soft tissues, its potential weaknesses should be further quantified. Because the PSTD software we have employed uses a regular, Cartesian grid to parameterize the medium of propagation, interfaces suffer from the “staircase effect”, resulting in spurious signals being generated near discontinuities\citep{van2002finite}.  We conducted three numerical experiments designed to visualize and quantify this effect via a simplified anatomy model, i.e., same as that of Fig.~\ref{fig:FIG3}, but with only one average biological soft tissue besides bone and water (Table~\ref{tab:table2}).
\begin{figure}[h]
\centering
\includegraphics[width=4.5in]{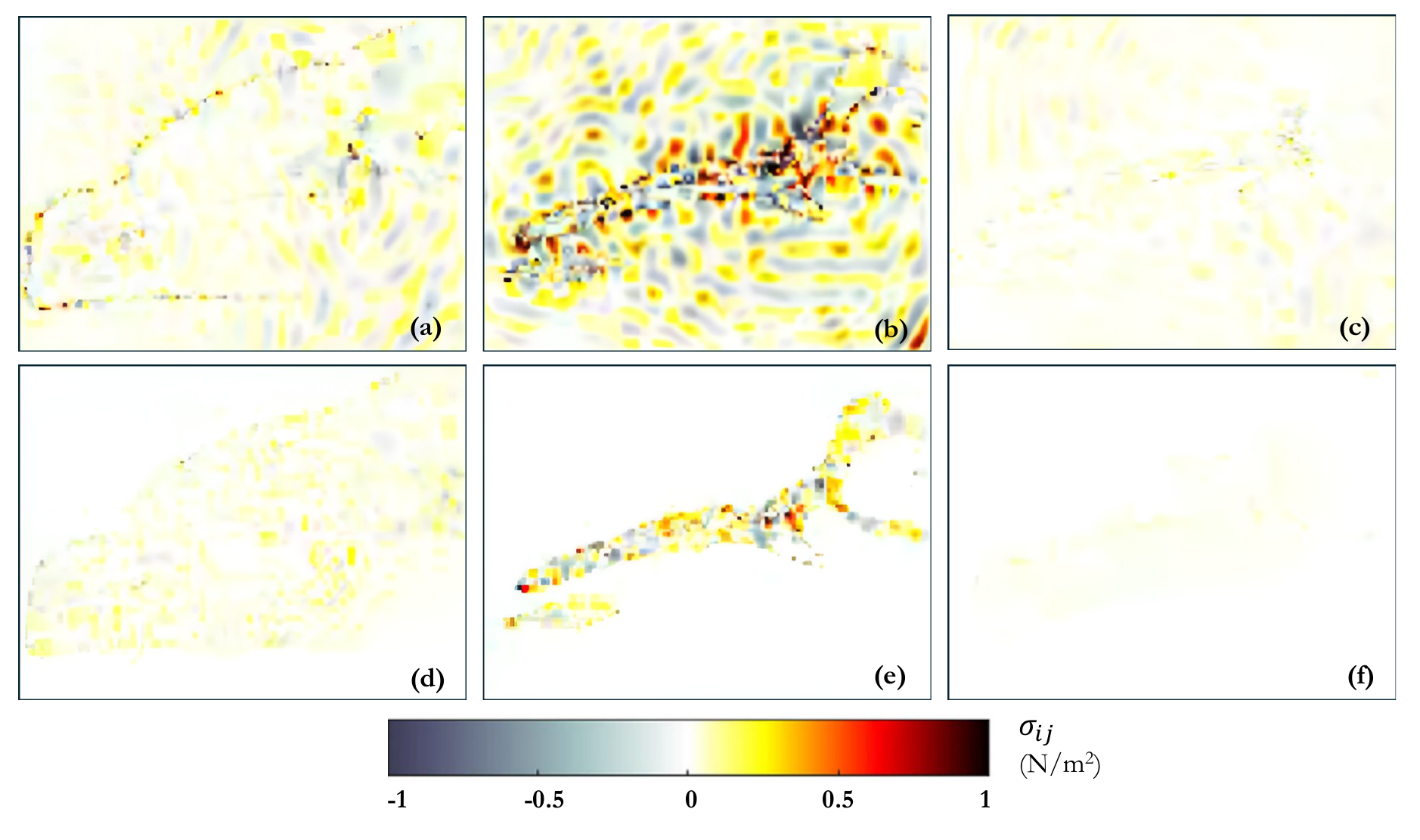}
\caption{\label{fig:FIG5}{Normal and shear stresses on the x–y plane of the reverberant field at simulation time, T = 0.5 ms. (a)–(c) Normal stress distributions for Case 1, Case 2, and Case 3, respectively. (d)–(f) Corresponding shear stress distributions for the same cases.}}
\end{figure}
\begin{table}[ht]
\caption{\label{tab:table2}Uniform elastic parameters for fluid--solid interface staircasing study.}
\renewcommand{\arraystretch}{1.3} 
\setlength{\tabcolsep}{10pt} 
\centering
\begin{tabular}{ccccccc}
\hline\hline
\text{Case}  &
\begin{tabular}[c]{@{}c@{}} Tissue \\ Type\end{tabular} & 
\begin{tabular}[c]{@{}c@{}}Averaged\\ Density ($\rho$)\\ {[}kg·m$^{-3}${]}\end{tabular} & 
\begin{tabular}[c]{@{}c@{}}Averaged\\ Compressional\\ Wave Speed ($c_p$)\\ {[}m·s$^{-1}${]}\end{tabular} & 
\begin{tabular}[c]{@{}c@{}}Averaged\\ Shear Wave Speed\\ {[}m·s$^{-1}${]}\end{tabular} \\ 
\hline
1 & Soft & 950 & 1400 & 1100 \\
  & Bone & 1963 & 3400 & 2500  \\
\hline
2 & Soft & 950 & 1400 & 0  \\
  & Bone & 1963 & 3400 & 2500 \\
\hline
3 & Soft & 950 & 1400 & 500 \\
  & Bone & 1963 & 3400 & 2500 \\
\hline\hline
\end{tabular}
\end{table}

Given the significant computational costs of our simulations, we limit ourselves to the 2-mm parameterization for this exercise. Our results in Fig.~\ref{fig:FIG5} show that the jump in shear wave speed at the fluid-solid (water-skin) interface essentially controls the extent of the staircase effect. Large artifacts in wave amplitude, sometimes saturating our color scale (and clearly violating the principle of energy conservation) appear near interfaces in Fig.~\ref{fig:FIG5}. In case 2 (in Figs.~\ref{fig:FIG5}b and ~\ref{fig:FIG5}e), where the shear wave speed contrast is about 2500~m$\cdot$s$^{-1}$, are more severe than in case 1, where shear wave speed contrast is only 1100~m$\cdot$s$^{-1}$ at the fluid-solid interface. When the shear wave speed in the tissue is reduced to 500~m$\cdot$s$^{-1}$ (case 3), artifacts at the discontinuities nearly disappear. To better understand this, we again employed a trial and error approach to tune our model. We determined that for complex structures similar to a dolphin's head, PSTD wave propagation models including all relevant tissues (bone, soft tissues, etc.) are stable, as long as the shear-wave speed ranges between 150~m$\cdot$s$^{-1}$ and 600~m$\cdot$s$^{-1}$. This constraint limits the numerical scheme's effectiveness in scenarios where the soft tissues in contact with high-density bones have shear-wave speed lower than 150~m$\cdot$s$^{-1}$. It should be noted that increasing the number of points per wavelength (PPW) can partially relax these constraints; however, this comes at the cost of significantly higher computational demand. A simulation test demonstrating this trade-off is discussed in Appendix A.
\section{\label{sec:3} RESULTS AND DISCUSSION}
\begin{figure}[!htbp]
\centering
\includegraphics[width=5.5in]{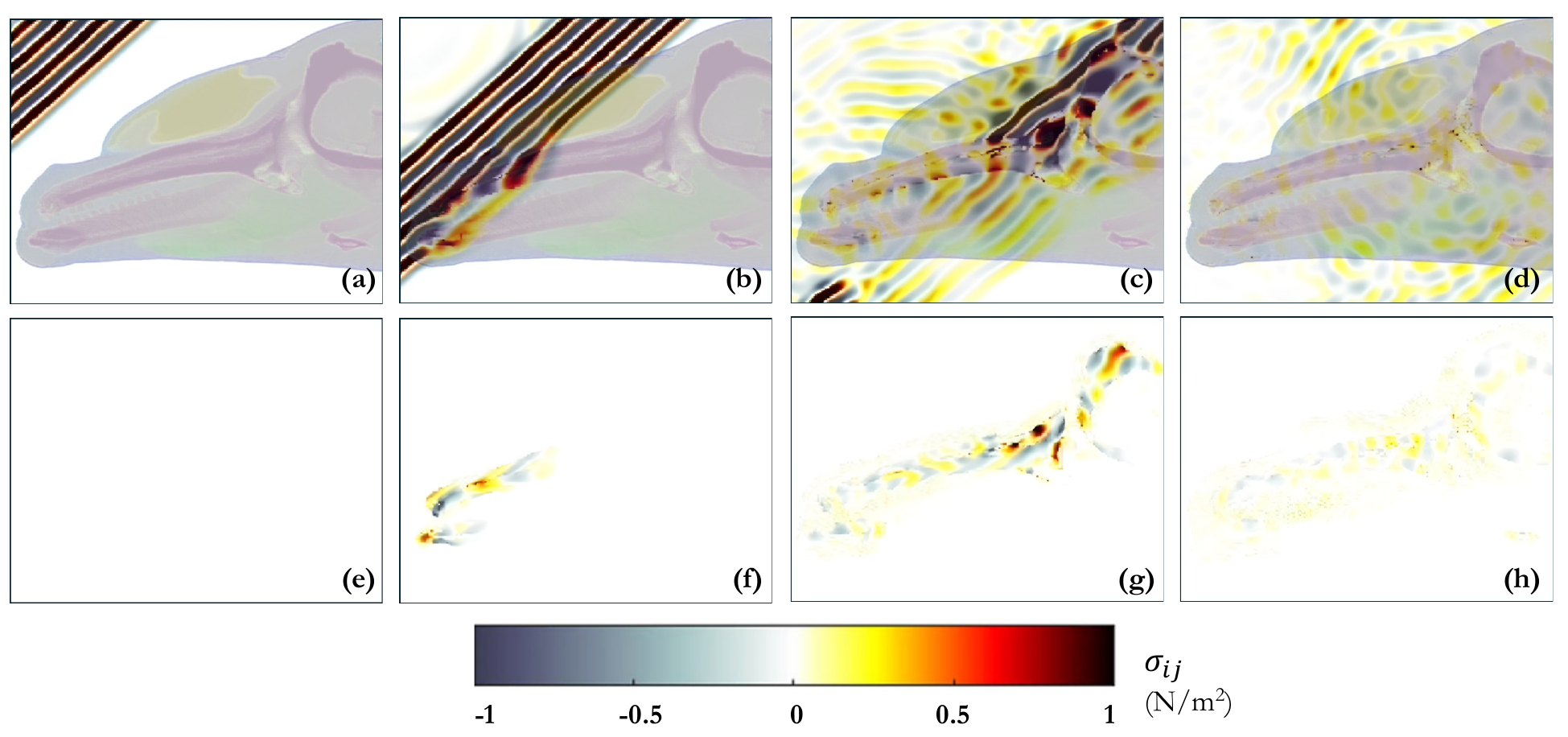}
\caption{\label{fig:FIG6}Normal and shear stress distributions on the x–y plane. (a)–(d) Normal stresses and (e)–(h) corresponding shear stresses showing snapshots of a $45^\circ$ plane wave sinusoidal burst at simulation times $T = 0.12$~ms (a), (e), $T = 0.20$~ms (b), (f), $T = 0.30$~ms (c), (g), and $T = 0.40$~ms (d), (h), respectively.}
\end{figure}
A suite of 37 simulations was conducted by initiating plane waves with propagation direction parallel to the median plane, and source-elevation angles spanning from $-90^\circ$ to $90^\circ$ in $5^\circ$ increments. Figure~\ref{fig:FIG6} illustrates the propagation of P-waves and S-waves at one selected elevation angle ($45^\circ$) along the dolphin head, where the time interval was set as $1\mathrm{e}{-7}$~s, with a sinusoidal burst as the signal source\citep{reinwald2018bone}. The Courant number\citep{courant1967partial} is set to $0.2748$, with a grid size of $1.11$ mm. Each simulation requires approximately 17 hours on 48 CPUs with 540 GB of RAM and involves 663 million degrees of freedom. 
\begin{figure}[!htbp]
\centering
\includegraphics[width=6in]{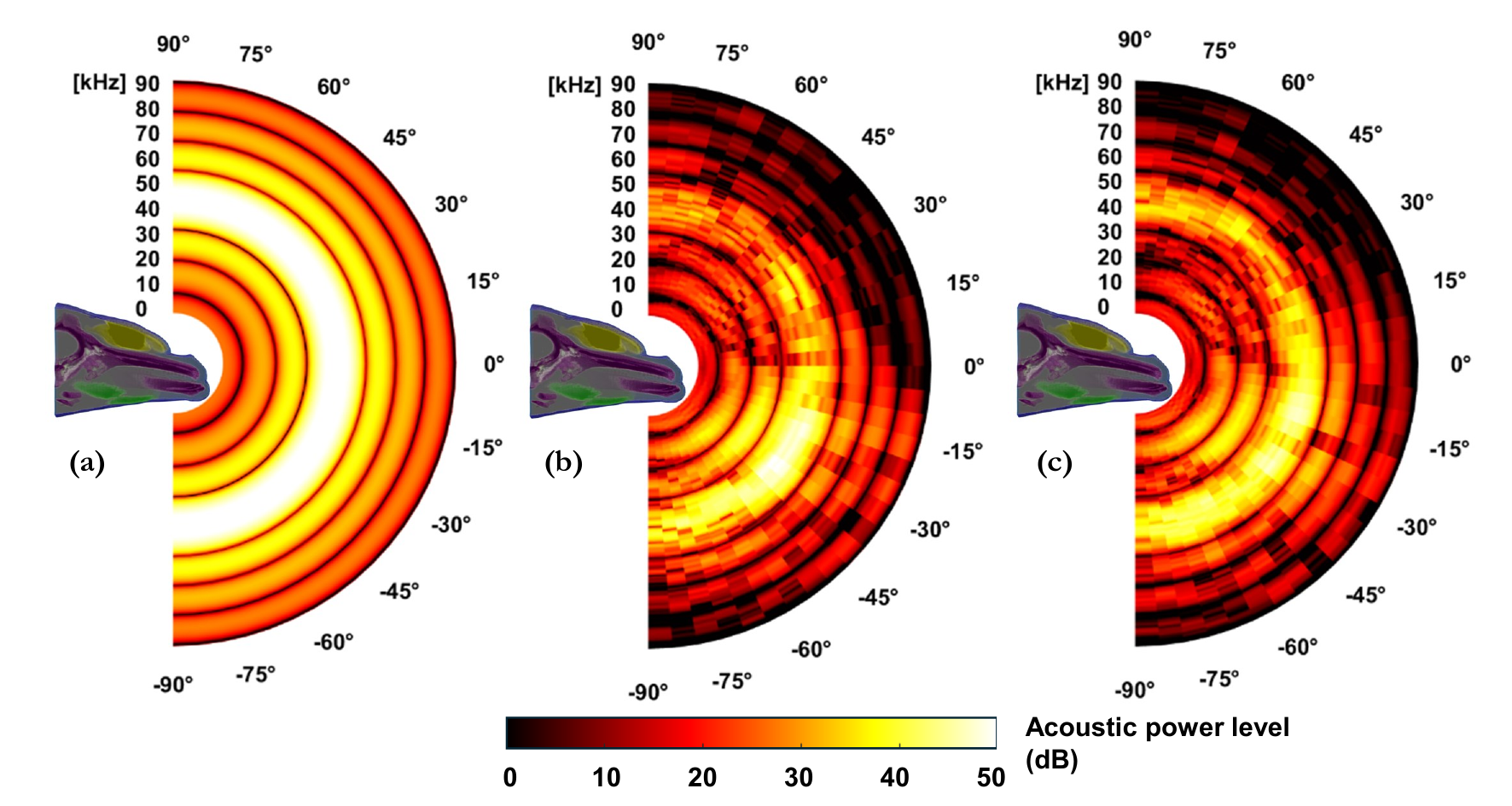}
\caption{\label{fig:FIG7}Frequency domain analysis of acoustic power spectra (in dB) for incoming sound directed from $-90^\circ$  to $+90^\circ$ : (a) input sinusoidal burst, (b) virtual sensor recordings at the left TPC, and (c) virtual sensor recordings at the right TPC.}
\end{figure}

The intensity level difference (ILD) and interaural time difference (ITD) computed at the TPC are below 2 dB and $4~\mu\mathrm{s}$, respectively: very small and consistent with previous studies\citep{reinwald2018bone, moore1995interaural}(see Appendix B). Further, the FFT of the recorded signals was performed for each elevation recorded at each left and right TPCs, as shown in Figs.~\ref{fig:FIG7}a–\ref{fig:FIG7}c. The results indicate that the amplitude of sound incoming at $+90^\circ$ to $+15^\circ$ is significantly reduced, particularly at higher frequencies. This is primarily because the TPC lies beneath the dorsal cranial structures, including the premaxillary and maxillary bones, which are directly exposed to incoming sound from higher elevation. Consequently, the entire dorsal surface reflects a substantial portion of high-frequency sound energy (60–90 kHz range) before it can reach the TPC. The lower jaw, on the other hand, is primarily exposed to incoming sound from $-15^\circ$ to $-60^\circ$. As a result, sound propagates along the lower jaw more efficiently. This result corroborates the theory of sound reception via the lower mandible, as previously documented\citep{norris1964some, norris1968evolution, brill2001assessment}.

\subsection{The transfer function at TPC}
\begin{figure}[!htbp]
\centering
\includegraphics[width=5.5in]{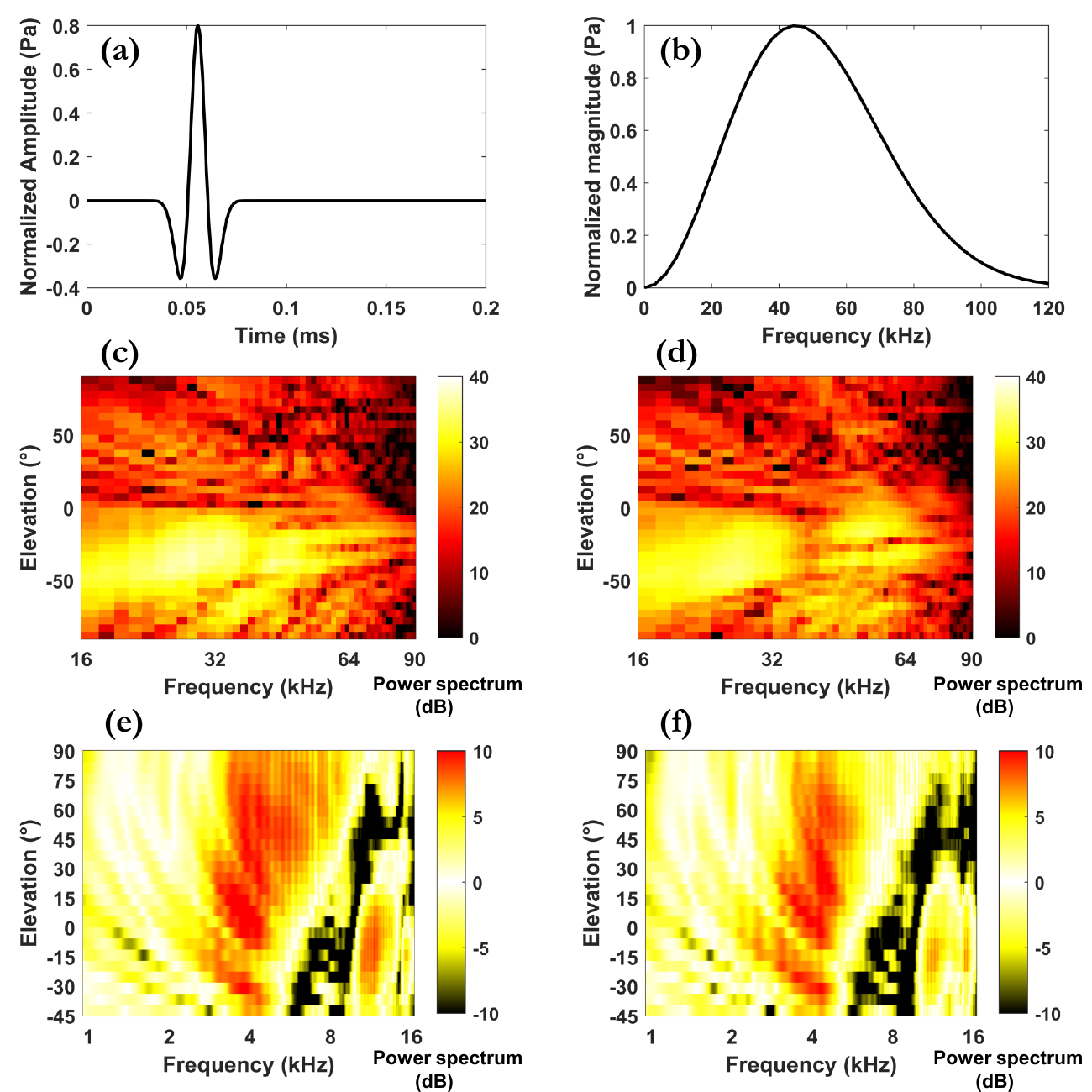}
\caption{\label{fig:FIG8}Analysis of TFs associated with the dolphin's head, focusing on the acoustic power spectrum (in dB). (a) Time-domain representation of a plane wave Ricker wavelet, (b) its corresponding frequency spectrum. (c) TF at the left TPC, (d) TF at the right TPC, (e) HRTF for the human left ear, and (f) HRTF for the human right ear. The human  HRTF we show is that of subject$-003$ from the CIPIC database\citep{algazi2001cipic}}
\end{figure}

The Transfer Function (TF), which is analogous to the HRTF in humans, is computed at the left and right TPCs as
\begin{equation}
H(f,\theta) = 20 \log_{10}  \mathcal{F}\big(|p(t,\theta)|\big).
\end{equation}
Where $H(f,\theta)$ denotes the frequency-domain energy spectrum in decibels (dB), and $p(t,\theta)$ the time-domain pressure recorded at the TPC for a given elevation $\theta$ on the median plane. The source time function is a Ricker pulse centered at 45 kHz, with relatively high amplitude between 32-64 kHz (Figs.~\ref{fig:FIG8}a and~\ref{fig:FIG8}b). The Ricker pulse has minimal energy content at lower frequencies and can therefore be handled more effectively by the PML compared to a Gaussian function. The resulting, approximated TFs are shown in Figs.~\ref{fig:FIG8}c and~\ref{fig:FIG8}d. While pronounced spectral notches are evident in the human HRTF within the 5–16 kHz range, particularly for elevation from  $-45^\circ$ to $+90^\circ$ along the median plane (in Figs.~\ref{fig:FIG8}e and~\ref{fig:FIG8}f), no distinct spectral notches between 32 and 64 kHz, are observed in the dolphin head.
\subsection{Biosonar simulation by correlation}
\begin{figure}[!htbp]
\centering
\includegraphics[width=5.5in]{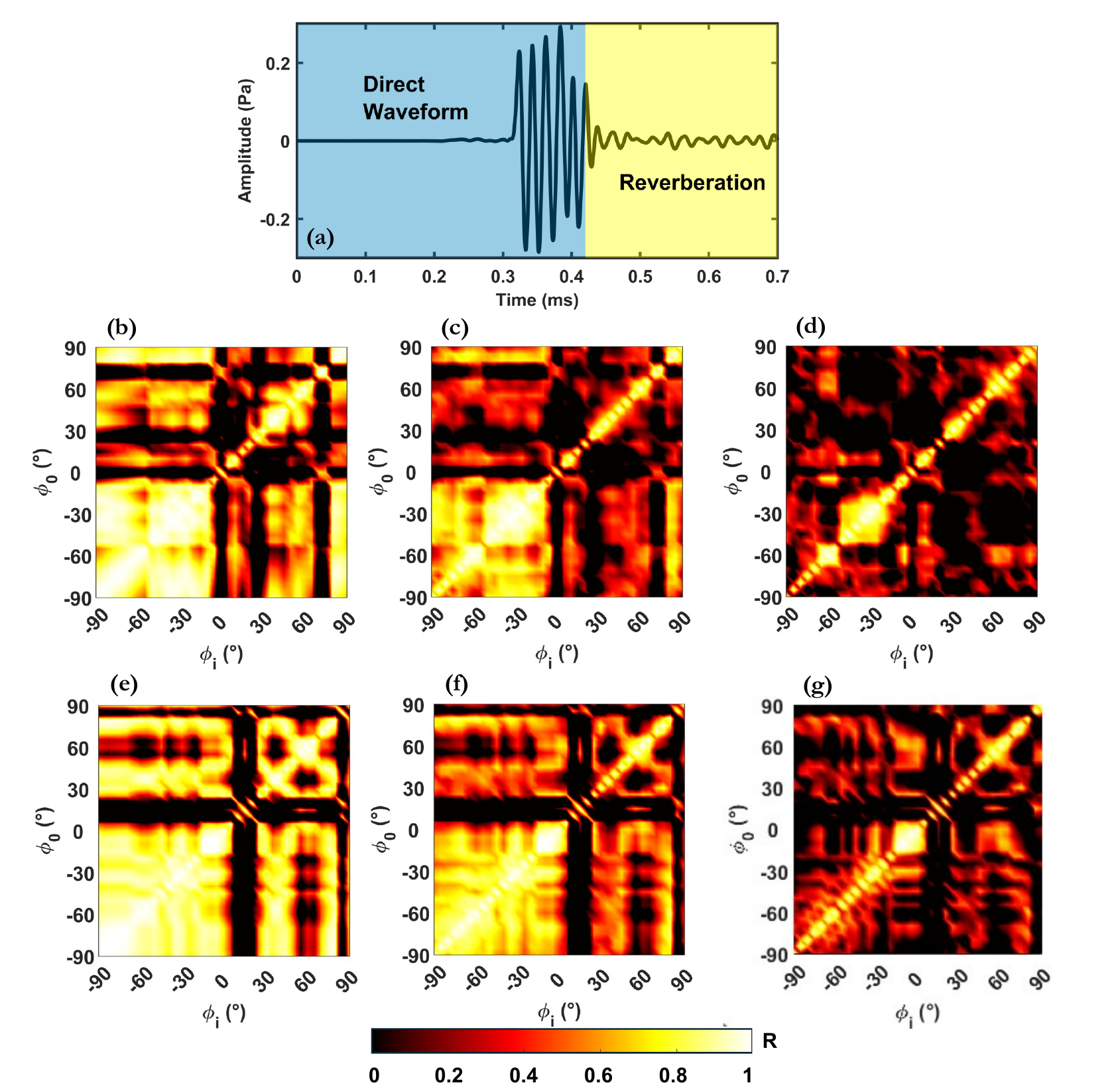}
\caption{\label{fig:FIG9}Sound localization by correlation: (a) Separation of the full waveform into direct (light blue) and reverberant (light yellow) components of the left TPC recording when a plane wave sinusoidal burst was fired at 0°. Pearson correlation coefficients of (b) the direct waveform, (c) the full waveform, and (d) the reverberant waveform at the left TPC. The corresponding results for the right TPC are shown in (e) the direct waveform, (f) the full waveform, and (g) the reverberant waveform, respectively.}
\end{figure}
Our simulations show that median-plane localization cannot be achieved via monaural spectral cues, cetaceans might take advantage of their own complex head structure through the reverberant field. To evaluate whether the TFs carries enough information to localize a source on the median plane, following Reinwald et al. (2018)\citep{reinwald2018bone}  and Hejazi Nooghabi et al. (2021)\citep{nooghabi2021contribution}  we compute the Pearson's correlation coefficient\citep{berman2016data} $R$ between a signal emitted from a given elevation and that recorded at each TPC, and the signals emitted at all other elevations and recorded at the same TPC,
\begin{equation}
R = \frac{\displaystyle \sum_t \bigl(p(\phi_0,t) - \overline{p_0}\bigr) \bigl(p(\phi_i,t) - \overline{p_i}\bigr)}
{\sqrt{\displaystyle \sum_t \bigl(p(\phi_0,t) - \overline{p_0}\bigr)^2} \; \sqrt{\displaystyle \sum_t \bigl(p(\phi_i,t) - \overline{p_i}\bigr)^2}} \;.
\label{eq:correlation}
\end{equation}
A perfect match between signals is achieved if $R = 1$, no match at all if $R = 0$, and a perfect anticorrelation (equal waveform, but of opposite sign) if $R = -1$. Here $\overline{p_0}$ denotes the mean value of the reference signal recorded at known elevation $\phi_0$, and $\overline{p_i}$ is the mean value of the signal at elevation $\phi_i$. All signals recorded at each TPC were first time-aligned using an algorithm that detects the first peak exceeding a threshold amplitude of $0.3$~Pa with reference to normalized waveform of 1 Pa amplitude. Each signal was then splitted into two components: the direct wave and the reverberation waveform. Pearsons’s $R$ values were computed for (i) the full waveform, (ii) the direct waveform, and (iii) the reverberated part of the waveform, separately for the left and right TPCs, as shown in Fig.~\ref{fig:FIG9}a.  As a rule, $R = 1$ on the diagonal of the matrices plotted in Figs.~\ref{fig:FIG9}b–\ref{fig:FIG9}g: we are correlating a signal with itself. Away from the diagonal, low correlation means that two signals can easily be distinguished from one another; conversely, the higher the correlation, the harder it is to discriminate source elevation. Including reverberated signal slightly improves this (compare Figs.~\ref{fig:FIG9}c and ~\ref{fig:FIG9}f with Fig. ~\ref{fig:FIG9}b and ~\ref{fig:FIG9}e). Correlating only the reverberated signal enhances  significantly the potential resolution of source elevation. These results suggest that the reverberant waveforms contain distinct and angle-specific information in the recordings from both the left and right TPCs (Figs.~\ref{fig:FIG9}d–\ref{fig:FIG9}g). These findings confirm previous experimental\citep{reinwald2018bone} and numerical results\citep{nooghabi2021contribution} on the skull of short-beaked dolphin, that were obtained neglecting the effects of soft tissues. 
\section{\label{sec:4} SUMMARY AND CONCLUSION}
In this study, we have tuned the k-Wave toolbox PSTD solver for the purpose of modeling compressional- and shear-wave propagation through marine-mammal anatomy. We confirm that the k-Wave package can capture many of the complexities of elastic wave propagation through intricate anatomical structures. We have next used it to investigate the dolphin biosonar, and its accuracy at discriminating the elevation of targets.  We have expanded earlier work by some of the authors of this study\citep{nooghabi2021contribution} to incorporate soft tissues in our model. We have addressed the question of whether this could lead to the emergence of distinct spectral notches in the TFs of dolphins, akin to those found in humans and bats. A principal summary follows. 
\begin{enumerate}[label=(\roman*)]
\item
	We have evaluated our model's grid resolution and staircasing phenomena, finding that grid resolutions of $1.11$~mm and 2 mm lead to very similar modeling results, suggesting that we are approaching convergence. In contrast, simulation results at 4 mm differ significantly.
\item
	We have found that, besides augmenting grid resolution, artifacts generated at discontinuities can be reduced by fine-tuning the values of model parameters. Specifically, values of shear-wave speed in soft tissues between 150 and 600~m$\cdot$s$^{-1}$ lead to stable results. This is fine, since our best guess for the value of shear velocity in tissues other than bones is close to 200~m$\cdot$s$^{-1}$.
\item
	The model was numerically validated using a spatial reciprocity test, in which the source and receiver were placed at different locations within the specimen’s head. Estimated relative numerical error is of the order of $10^{-4}$.
\item
	No distinct spectral notches were observed in the TFs of the specimen’s head at each TP complex. On the other hand, Pearson's correlation coefficient indicates that the dolphin's head can function in principle like an antenna, providing enough information to fully discriminate median-plane sources from one another, with accuracy of at least $5^\circ$. 
\end{enumerate}

The computational performance of the current PSTD method was also evaluated. An additional test was carried out at higher resolution than used throughout this paper, on the simplified water/bone model of Table II, case 2. The original CT scan $(0.813 \times 0.813 \times 1.5~\text{mm})$ was interpolated to a $0.5$ mm grid. Using the same CPUs configuration as above (i.e., 48 CPUs with 540 GB of RAM), one simulation required approximately 280 hours (around 16 times more than the $1.11$~mm resolution simulations discussed above). It remains to be determined whether this figure can be reduced significantly, upon migrating  our software to a high-performance computer cluster. For now, we submit that the PSTD approach can potentially be extended to higher-resolution simulations than achieved here. This will be critically important for future investigations, based on high-resolution CT scans and especially for detailed numerical modeling of  high-frequency sound propagation through small-scale structures such as the TPC.  

\section*{Acknowledgment}
This project has received fundings from the MSCA Doctoral Network SEASOUNDS (Innovative marine soundscape characterization to effectively mitigate ocean and sea noise pollution, Grant agreement no. 101119769), coordinated by CNRS (France). We gratefully acknowledge Jean-Marie Graïc (Department of Comparative Biomedicine and Food Science, University of Padova) for providing the CT scan data. We also thank Maxence Ferrari, Paul Cristini, Nathalie Favretto-Cristini, Vadim Monteilier (CNRS, Laboratory of Mechanics and Acoustics), and Jakob Tougaard (Department of Ecoscience, Aarhus University) for their insightful comments.


\section*{APPENDIX A:  THE STAIRCASING EFFECT}
\begin{figure}[h]
\centering
\includegraphics[width=4.5in]{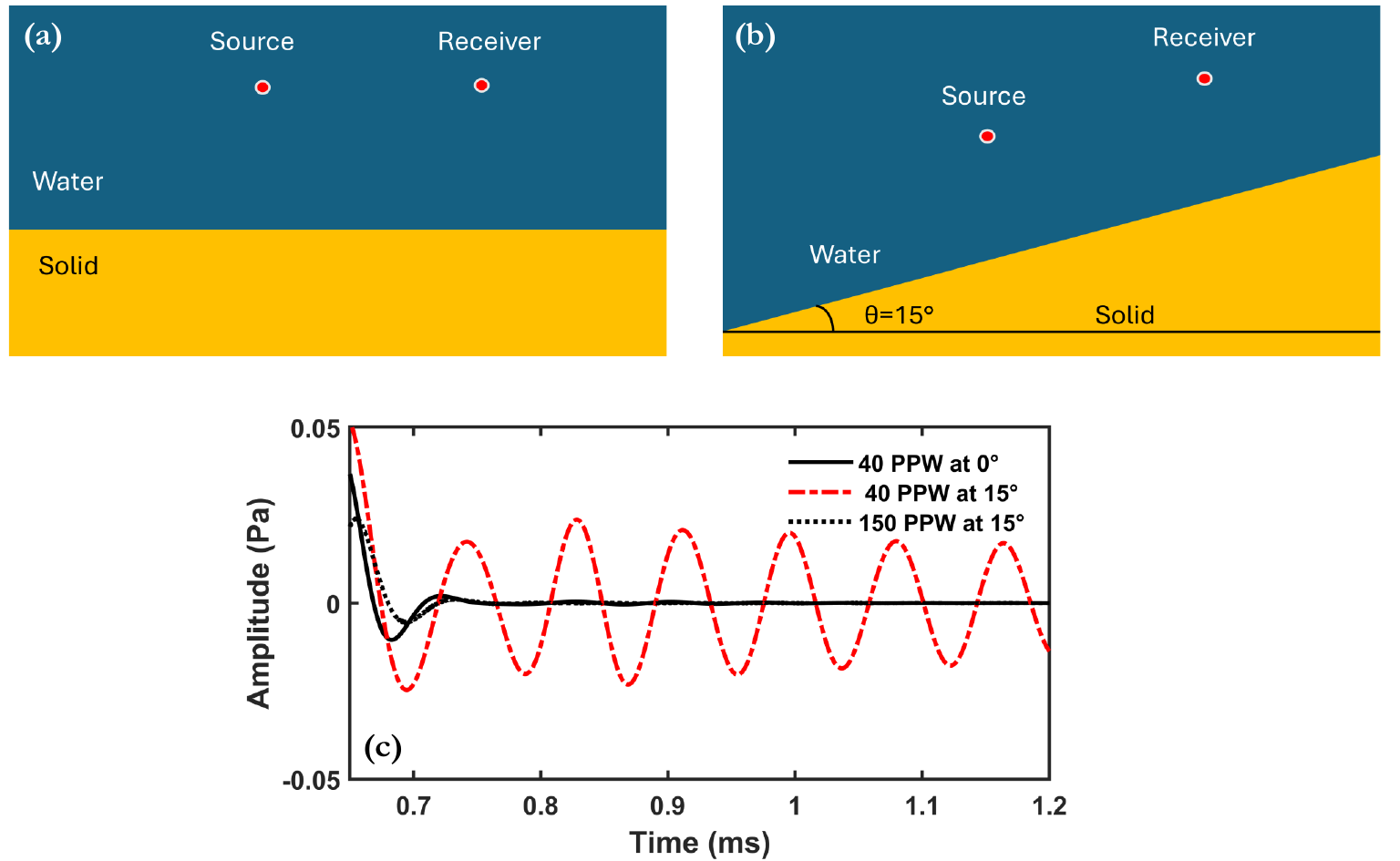}
\caption{\label{fig:FIG10}Staircasing effects demonstrated at 40 and 150 parts per wavelength (PPW). 3D domain subjected to (a) 0° and (b) 15° tilt, following the setup of Firouzi et al. (2012\citep{firouzi2012first}). (c) Reflected waveform from the solid surface measure at the receiver point for both (a) and (b) configurations.}
\end{figure}
A simple 3D numerical model has been implemented, consisting of two layered media, i.e., solid ($c_\mathrm{s} = 2500\ \mathrm{m\,s^{-1}}$, $c_\mathrm{p} = 3400\ \mathrm{m\,s^{-1}}$, and $\rho = 1963\ \mathrm{kg\,m^{-3}}$) and water ($c_\mathrm{s} = 0\ \mathrm{m\,s^{-1}}$, $c_\mathrm{p} = 1500\ \mathrm{m\,s^{-1}}$, and $\rho = 1000\ \mathrm{kg\,m^{-3}}$), separated by a discontinuity and bounded by PML absorbing boundaries. A Ricker point pulse centered at 4 Hz source was deployed 870 m above the solid surface and 600 m away from the receiver point  (Fig.~\ref{fig:FIG10}a). We next repeated the exact simulation after rotating the reference frame, and, therefore, the numerical grid, by 15° as per Fig.~\ref{fig:FIG10}b.  The difference between modeled signal at the receiver is a purely numerical artifact resulting from the "pixelization" of the interface, and often referred to as ``staircase effect''.  In a first simulation, with grid resolution at 40 PPW, the error is severe. At 150 PPW it has all but disappeared (Fig.~\ref{fig:FIG10}c).
\section*{APPENDIX B: INTERAURAL LEVEL AND TIME DIFFERENCES}
\begin{figure}[h]
\centering
\includegraphics[width=5in]{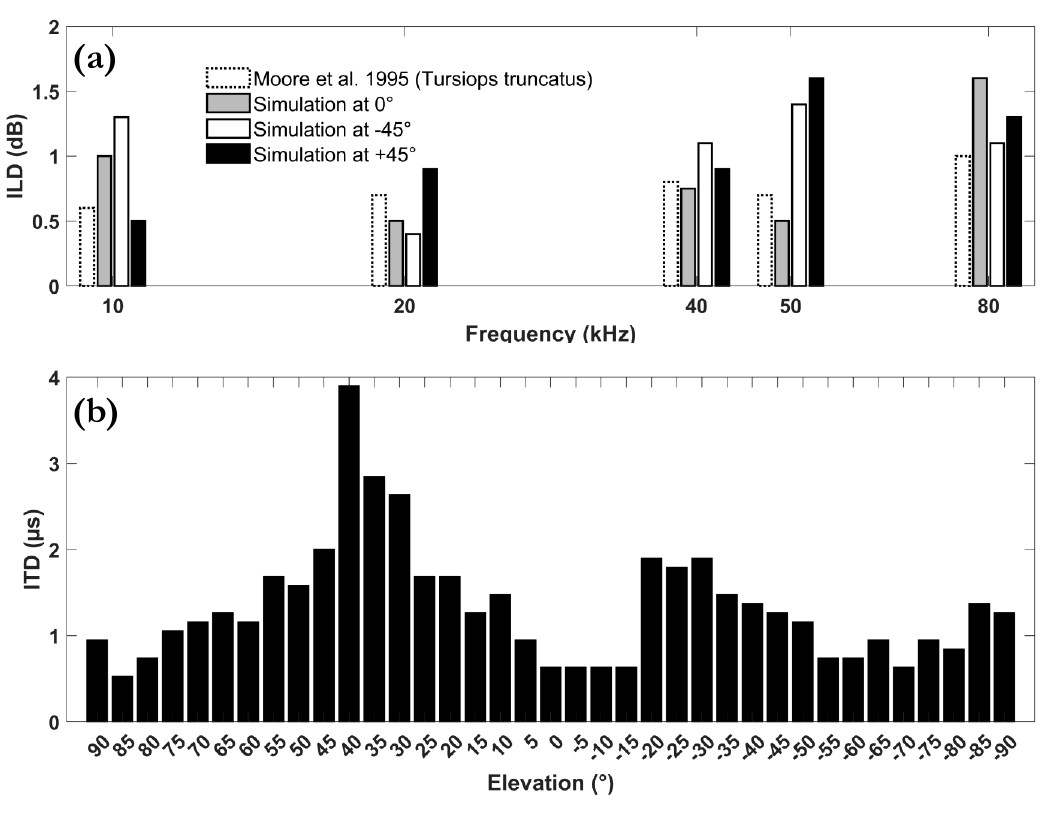}
\caption{\label{fig:FIG11}Comparison of the (a) intensity level difference (ILD) at various frequencies, and (b) the onset interaural time difference (ITD) as a function of median-plane source elevation.}
\end{figure}
Figure.~\ref{fig:FIG11}, shows the ILD and ITD obtained from sources within the median plane and receivers at the approximate TPC locations. They are generally very small, compared to typical values  e.g., $\pm 18~\,\mathrm{dB}$ (ILD) and about $-70$ to $90~\,\mu\mathrm{s}$ (ITD) for sources located on the horizontal plane at azimuths from -90° to 90°, respectively\citep{reinwald2018bone, moore1995interaural}.

\bibliographystyle{IEEEtran}  
\bibliography{Main-bib}

\end{document}